# Crystals, magnetic and electronic properties of a new ThCr$_2$Si$_2$-type BaMn$_2$Bi$_2$ and K-doped compositions

Bayrammurad Saparov*, Athena S. Sefat

Materials Science & Technology Division, Oak Ridge National Laboratory, P. O. Box 2008, Building 4100, 1 Bethel Valley Road, Oak Ridge, TN 37831-6056, USA.

E-mail: saparovbi@ornl.gov

**Abstract.** This is a report on the new 122 ternary transition-metal pnictide of BaMn$_2$Bi$_2$, which is crystallized from bismuth flux. BaMn$_2$Bi$_2$ adopts ThCr$_2$Si$_2$-type structure (*I*4*/mmm*) with $a$ = 4.4902(3) Å and $c$ = 14.687(1) Å; it is antiferromagnetic with anisotropic magnetic susceptibility and semiconducting with a band gap of $E_g$ = 6 meV. Heat capacity result confirms the insulating ground state in BaMn$_2$Bi$_2$ with the electronic residual Sommerfeld coefficient of $\gamma \approx 0$. The high temperature magnetization results show that magnetic ordering temperature is T$_N$ ~ 400 K. Hole-doping in BaMn$_2$Bi$_2$ via potassium in Ba$_{1-x}$K$_x$Mn$_2$Bi$_2$ results in metallic behavior for $x$ = 0.10(1), 0.32(1) and 0.36(1). With K-doping, more magnetically anisotropic behavior is observed. Although there is a downturn in electrical resistivity and low-field magnetization data below 4 K in > 30%-doped crystals, there is no sign of zero resistance or diamagnetism. This manuscript is a report on new materials of Ba$_{1-x}$K$_x$Mn$_2$Bi$_2$ (0 ≤ $x$ < 0.4). Results from powder X-ray diffraction, anisotropic temperature- and field-dependent magnetization, temperature-and field-dependent electrical resistivity, and heat capacity are presented.

## Introduction

Materials with ThCr$_2$Si$_2$-type structure, "122", have attracted interest of solid state community for decades for the diverse bonding patterns and dimensionality that give rise to exotic physical properties. [1-6] The changes in bonding patterns and dimensionality result in interesting effects, for example, in SrCo$_2$(Ge$_{1-x}$P$_x$)$_2$, a ferromagnetic quantum critical point is induced by interlayer anion dimer breaking.[3] Also in Ba$_{1-x}$K$_x$Fe$_2$As$_2$ and BaFe$_{2-x}$Co$_x$As$_2$, unconventional superconductivity was recently uncovered.[4-6] The discovery of superconductivity in these iron-based arsenide quasi-2D ThCr$_2$Si$_2$-type structures[5] with superconducting transition temperatures, $T_C$, up to ≈ 45 K, has led the research path to finding varieties of physical properties in other 3$d$-based compounds such as itinerant antiferromagnetism in $BaCr_2As_2$,[7] insulating antiferromagnetism in $BaMn_2As_2$,[8,9] metallic diamagnetism in $BaCu_2As_2$,[10] correlated metallic behavior in $BaCo_2As_2$,[11] and superconductivity in $BaNi_2As_2$ ($T_C$ ≈ 0.6 K).[12] Extensive characterizations of these arsenides has been limited due to the unavailability of high quality crystals for the toxicity and the high melting points of binary arsenides $TM$As ($TM$ = transition metal) that are used as fluxes for growing crystals.[6,12]

In order to understand the causes of high-temperature superconductivity, investigations on new 122 classes of non-Fe-based superconductors would be crucial. Additionally, because of the toxic nature of arsenic, there is an incentive for chemical substitutions with other pnictogens. Moreover, the heavier pnictides of antimonides and bismuthides may have a higher $T_C$ compared to the smaller arsenide and phosphide analogs: for example LaFeAsO has a higher $T_C$ compared to LaFePO, and LiFeSb is predicted[13] to have a higher $T_C$ compared to LiFeAs.[14] In this study, we have done exploratory synthesis aimed at stabilizing a transition-metal 122 bismuthide. In fact, there are no bismuthides with $ThCr_2Si_2$ structure in the Inorganic Crystal Structure Database (ICSD).[15] In our quest to synthesize and study the first bismuthide with $ThCr_2Si_2$ structure, we singled out $BaMn_2Bi_2$ as a target, partly because $BaMn_2Sb_2$[16] exists as the only known barium antimonide among 3$d$ transition metals, and therefore, it was assumed that $BaMn_2Bi_2$ is more likely to form.

A survey of the reports on barium manganese pnictides reveals that $BaMn_2P_2$ exhibits $G$-type antiferromagnetism, where moments are aligned AF along all crystal axes, with Neel ordering temperature ($T_N$) above 750 K,[17] and it is a semiconductor with a band gap of $E_g$ = 146 meV. $BaMn_2As_2$ is also a $G$-type antiferromagnet below $T_N$ = 625 K with anisotropic magnetism and a small band gap of $E_g$ = 54 meV.[8,9,18] Investigations of infrared and Raman optical properties suggest that $BaMn_2As_2$ is much more two dimensional in its electronic properties compared to the parent Fe-based superconductor of $BaFe_2As_2$.[19] Under high pressures, $BaMn_2As_2$ exhibits metallic behavior with a downturn in resistivity data below ~17 K.[20] Hole doping via substitution of Ba with K results in metallic $Ba_{1-x}K_xMn_2As_2$ ($x$<0.06)[21] samples. The crystal structure is preserved, whereas the ordering temperature reduces to $T_N$ = 607 K for 5% doping and for higher doping levels of 19% and 26%, weak ferromagnetism is seen below ~50 K.[22] Theoretical and experimental work suggests that $BaMn_2Sb_2$ also exhibits high temperature antiferromagnetic *G*-type ordering.[9,23] Measurements of transport properties reveal that BaMn$_2$Sb$_2$ has a room temperature Seebeck coefficient of ~225 µV/K, but the electrical conductivity is low resulting in poor thermoelectric properties.[24] Band gap calculated from the resistivity data is estimated to lie in the ranges of $E_g$ = 68-204 meV.

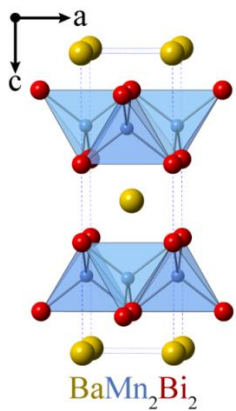

**Fig. 1** Polyhedral representation of the crystal structure of BaMn$_2$Bi$_2$ (ThCr$_2$Si$_2$-type). The polyanionic [Mn$_2$Bi$_2$]$^{2-}$ layers are emphasized, and a unit cell is outlined.

In this work, we report synthesis of large single crystals of BaMn$_2$Bi$_2$, the first bismuthide crystallizing in ThCr$_2$Si$_2$-type structure (Fig. 1). The crystals are grown from excess Bi ($T_{mp}$ = 271°C). We discuss the results of powder X-ray diffraction, energy-dispersive X-ray spectroscopy, temperature- and field-dependent magnetization, temperature-dependent electrical resistivity and heat capacity, and differential thermal analysis. Effects of K-hole doping are also reported through single crystal synthesis and physical property measurements of Ba$_{1-x}$K$_x$Mn$_2$Bi$_2$ for $x$ = 0.1, 0.32 and 0.36.

**Experimental**

**Synthesis**

Elemental reactants of dendritic Ba (99.9%), Mn pieces (99.95%), Bi granules (99.9999%), and K (99.95%) were used as received from Alfa Aesar.

Single crystals of BaMn$_2$Bi$_2$ were grown inside an alumina crucible, in an excess of bismuth using a Ba:Mn:Bi = 1:2:10 or 1:2:5, with the latter ratio providing larger crystals. The crucible containing reactants was sealed in vacuum inside a silica tube. The reaction mixture was placed inside a box furnace and heated to 1000°C at a rate of 200°C h$^{-1}$, annealed 15 h at this temperature, then cooled to 415°C at a rate of 5°C h$^{-1}$. At this temperature, the reaction setup was centrifuged to remove excess liquid bismuth. Large plate-shaped crystals of BaMn$_2$Bi$_2$ measuring up to 1 cm in length (see inset of Fig. 2) were the only products of the reaction. Magnetization results, however, indicated presence of small (0.24% from magnetization results) amount of MnBi impurity coating crystals. MnBi-free smaller crystals (1 to 2 mm) were grown by reaction of elements in a Ba:Mn:Bi = 1:2:8 ratio using the same heating profile, but removing the excess flux at 500°C. Although BaMn$_2$Bi$_2$ is stable under air for several hours, the crystal surface becomes dull with overnight exposure to air.

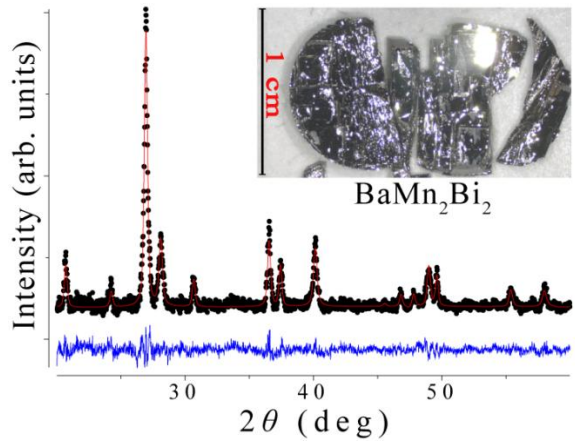

**Fig. 2** Powder X-ray diffraction pattern for BaMn$_2$Bi$_2$ (black dots). The Rietveld fit and the difference plot are shown in red and blue, respectively. The refined unit cell parameters are $a$ = 4.4902(3) Å and $c$ = 14.687(1) Å, and $z_{Bi}$ = 0.3692(3). The inset shows the typical crystal size.

The reaction conditions for K-doped single crystals, Ba$_{1-x}$K$_x$Mn$_2$Bi$_2$, had to be modified due to high volatility of K at relatively low temperatures (T$_{bp}$ = 760°C) and its reactivity with silica in the vapor phase. Reactions with low K content, nominal $x$ = 0.4, were synthesized by heating the reaction mixture to 900°C, dwelling at this temperature for 15 h, and then cooling to 415°C at a rate of 4°C h$^{-1}$, followed by the removal of the bismuth flux. Crystal quality was comparable to that of the parent compound, and no secondary phase was detected. Higher K content reactions required further modification of the synthesis conditions: for doping of nominal

$x = 0.6$ and $x = 0.8$, in addition to the sealing the reactions under a 2/3 atm of ultra-high purity Ar, they were only heated up to 700°C and kept for 8 h, followed by cooling to 415°C at a rate of 4°C h$^{-1}$, and then centrifugation. The products contained a mixture of plate and needle crystals. The relative ratio of needle- to plate-shaped crystals increased with K content, suggesting that the secondary phase is K-rich. No traces of MnBi impurity is observed in magnetization data of doped crystals.

**Characterization**

*Elemental analysis*

Scanning electron microscope imaging and energy-dispersive X-ray spectroscopy (EDS) were carried out using a Hitachi-TM3000 microscope equipped with a Bruker Quantax 70 EDS system. Data acquisition was carried out with an accelerating voltage of 15 kV in 2-3 min scanning time. Samples with high K-doping were air (probably oxygen and moisture) sensitive. The samples with nominal K contents of $x = 0.4$, 0.6 and 0.8 gave EDS content of $x = 0.10(1)$, 0.32(1) and 0.36(1), respectively. The samples are denoted by these values throughout the paper. The elemental composition of the secondary needle crystals could not be determined unequivocally as the composition changed from crystal to crystal, and in some cases the same crystal showed different compositions in various spots, however, the most recurring content was K:Mn:Bi ≈ 1:5:4.

*X-ray diffraction*

Room temperature powder X-ray diffraction (XRD) patterns were collected on a PANalytical X'Pert PRO MPD X-ray diffractometer. An incident beam Johansson monochromator was used to get Cu-K$\alpha_1$ radiation. A typical run was performed in the 2$\theta$ range of 5 to 65° with a step size of 1/60° and 40 seconds/step counting time in a continuous scanning mode. Upon grinding, crystals were found to be malleable; the ground sample burned in air, therefore special care was taken for the preparation for XRD, which consisted of grinding crystals inside a helium-filled glovebox, sealing them in a special air-tight holder with polycarbonate cover, and then transferring them onto the goniometer. Although the polycarbonate cover increased the background intensity, the diffraction pattern of BaMn$_2$Bi$_2$ was suitable for the Rietveld refinement of the unit cell parameters and the atomic coordinates (see Fig. 2) using GSAS[25,26] software package. In addition, Le Bail fits were successfully applied to XRD patterns of Ba$_{1-x}$K$_x$Mn$_2$Bi$_2$ for doping levels $x = 0.1$ and $x = 0.32$. However, $x = 0.36$ could only be indexed qualitatively due to the small yield.

*Physical property measurements*

Magnetization measurements were carried out using a Quantum Design Magnetic Property Measurement System (MPMS). The temperature dependence of magnetization experiments were performed on single crystals under an applied field of 10 kOe along the *c*-axis and *ab*-plane. Both zero field cooled (ZFC) and field cooled (FC) data were collected below 350 K. Measurements were also performed at 20 Oe, below 30 K, to check for a possible diamagnetic signal. Field-dependent magnetization measurements were performed at 5 K and 295 K. For the air-sensitive crystals, they were covered with a small amount of superglue in the glovebox before transferring to MPMS. High temperature magnetization measurements on $BaMn_2Bi_2$ were carried out using quartz tubings with an inner diameter of ~ 1.75 mm. Several small crystals were put into quartz tubings, and approximately oriented along the *c*-axis and *ab*-plane inside the glovebox. The tubings were enclosed using quartz rods of fitting size, then quickly taken out of the glovebox and flame-sealed. Temperature dependent magnetic susceptibility measurements were carried out from 310 K to 780 K under 10 kOe. Field-dependent magnetization measurements were also performed at 780 K.

Four-probe electrical resistivity and heat capacity measurements were carried out on a Quantum Design Physical Property Measurement System (PPMS). Electrical contacts were made using platinum wires and Dupont 4929N silver paste. The contacts placement and transfer to the PPMS took less than 3 min for crystals with $x$ = 0.1, 0.32 and 0.36. Typical measurements were in the 1.8 to 400 K range under zero and 50-80 kOe applied fields. Heat capacity measurement was only carried out for $BaMn_2Bi_2$ below 200 K using the relaxation method. The $C/T$ vs $T^2$ plot was fitted at low temperature region to extract the lattice ($\beta$) and electronic ($\gamma$) contributions. The Debye temperature ($\theta_D$) was calculated using the relationship $\beta = 12\pi^4 NR/5\theta_D^3$, where $R$ = 8.314 J/(K·mol).

*Thermal analysis*

Thermogravimetric analysis (TGA) and differential thermal analysis (DTA) of $BaMn_2Bi_2$ were performed using a Pyris Diamond TG/DTA from Perkin Elmer under a stream of ultra-high purity argon gas. The measurement was carried out on a 36.5 mg plate crystal in the temperature interval 318-790 K with a heating rate of 30 K min$^{-1}$ and a cooling rate of 10 K min$^{-1}$.

**Results and discussion**

**Crystal structure, chemistry and stability**

Fig. 2 (inset) shows BaMn$_2$Bi$_2$ crystals from a reaction with a total mass 4.3 g. The crystal growth was carried out inside a 2 ml alumina crucible with Bi flux, which resulted in crystals measuring up to 1 cm in length. Such a reaction is likely to afford growth of larger crystals, if the reaction is scaled up with larger crucibles.

BaMn$_2$Bi$_2$ crystallizes in the body centered tetragonal space group *I*4/*mmm* (No. 139), ThCr$_2$Si$_2$-type structure, Pearson symbol *tI*10. The BaMn$_2$Bi$_2$ structure is built of [Mn$_2$Bi$_2$]$^{2-}$ polyanionic layers separated by Ba$^{2+}$ along the *c*-axis. The [Mn$_2$Bi$_2$]$^{2-}$ layers, in turn, are made of edge-sharing MnBi$_4$ tetrahedra fused into a PbO-type layer (Fig. 1). The compound is isostructural with the lighter pnictide analogs BaMn$_2$P$_2$, BaMn$_2$As$_2$ and BaMn$_2$Sb$_2$. Detailed description of ThCr$_2$Si$_2$-type structure, and its relationship with other structures, are already reported.[1,27]

Fig. 2 shows the powder XRD pattern of BaMn$_2$Bi$_2$. The low angle 2θ region of 5-20° was omitted because of no structural Bragg peaks and a broad background caused by the polycarbonate cover. The broadening of the Bragg peaks is due to malleability of this compound; this feature is also reported for BaMn$_2$As$_2$[9] and CaFe$_2$As$_2$.[28] The Rietveld refinement of the diffraction data gives *a* = 4.4902(3) Å and *c* = 14.687(1) Å, with a reliability factor *Rp* = 7.53 % (weighted *R*-factor *wRp* = 9.92 %), and goodness-of-fit $\chi^2$ = 1.213. The *z* coordinate for Bi is refined to $z_{Bi}$ = 0.3692(3). The soft and ductile nature of BaMn$_2$Bi$_2$ resulted in smeared rings in single crystal X-ray diffraction, which prevented the collection of quality data and full refinement, although agreeable unit cell parameters were identified.

The unit cell dimensions for BaMn$_2$Bi$_2$ are larger compared to the lighter pnictides BaMn$_2$P$_2$, BaMn$_2$As$_2$, and BaMn$_2$Sb$_2$, and in fact, it has the largest unit cell volume among all pnictides with ThCr$_2$Si$_2$-type structure. The Mn-Bi bond distance in the [Mn$_2$Bi$_2$]$^{2-}$ layer is $d_{Mn-Bi}$ = 2.847(2) Å, and the tetrahedral angles are 104.1(1)° and 112.22(6) °. The interlayer distance shown by the distance between two bismuth atoms in adjacent layers is $d_{Bi-Bi}$ = 3.842(8) Å, whereas the distance between Ba and Bi is $d_{Ba-Bi}$ = 3.711(2) Å. BaMn$_2$Bi$_2$ is one of only two known compounds in the ternary Ba-Mn-Bi system.[15,29] The other known compound, Ba$_{14}$MnBi$_{11}$,[30] also features tetrahedral MnBi$_4$ building blocks with the $d_{Mn-Bi}$ = 2.935(1) Å bond distance and tetrahedral angles of 104.5(1) ° and 119.9(1)°. Tetrahedra are almost ideal in the related KMnBi,[31] which is isostructural with the superconducting LiFeAs (PbClF/Cu$_2$Sb-type structure), with angles 109.4-109.7° and the distance $d_{Mn-Bi}$ = 2.900.

The BaMn$_2$Bi$_2$ crystals are stable under ambient air for brief periods, but if left out overnight, the surface of the plates become dull and grainy. Finely ground BaMn$_2$Bi$_2$ powder burns in air instantaneously, producing Bi and BaBi$_3$ as byproducts (Fig. 3). BaBi$_3$ is also air sensitive.[32] We confirmed this by doing XRD runs after 30 min exposure to air, which shows a considerable decrease in the BaBi$_3$ amount, and finally after overnight exposure, no BaBi$_3$ was present. Powder XRD data of a sample left overnight in air shows only peaks from elemental Bi, and an additional peak attributed to MnO$_2$.

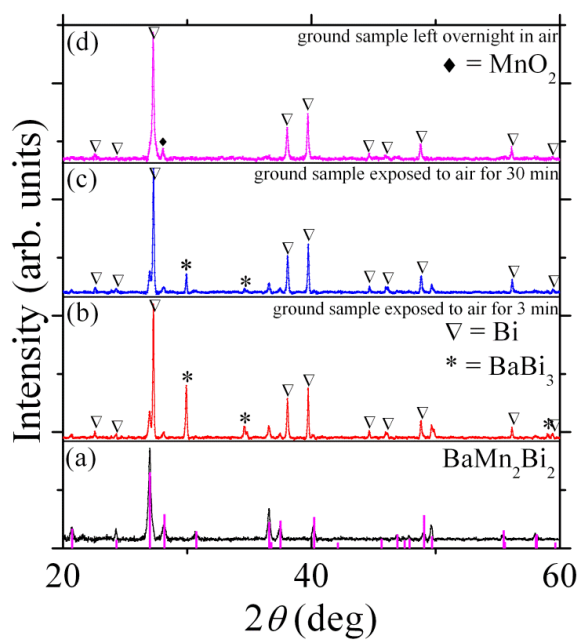

**Fig. 3** Powder X-ray diffraction pattern for (a) BaMn$_2$Bi$_2$ (black line) matched with the calculated pattern (pink tick marks). (b) BaMn$_2$Bi$_2$ after 3 min exposure to air, resulting in a crystalline mixture of BaMn$_2$Bi$_2$, Bi and BaBi$_3$. (c) BaMn$_2$Bi$_2$ after 30 min exposure to air with crystalline BaMn$_2$Bi$_2$ and BaBi$_3$ contents further reduced. (d) BaMn$_2$Bi$_2$ left overnight in air, showing Bi and MnO$_2$ crystalline phases.

The DTA data for BaMn$_2$Bi$_2$ show (Fig. 4) one endothermic effect on heating, and one exothermic effect on cooling; these features seem to correspond to melting and freezing point of elemental Bi, respectively. The source of Bi is likely due to the crystal surface oxidation (see Fig. 3 for $BaMn_2Bi_2$ oxidation byproducts), in addition to a small coating of Bi on crystal surface. Although clear signs of the surface oxidation were visible after the experiments, the inner parts of the crystal preserved its characteristic metallic luster. The relative mass change was less than 0.2% during this experiment according to the TGA data. In order to prove that the effects seen on the DTA data are due to the presence of Bi produced by surface oxidation, several crystals of $BaMn_2Bi_2$ were placed inside an aluminum crucible, which was then vacuum sealed inside a silica tube. The container was heated to 573 K in 100 K intervals. The surface of the crystals remained fairly shiny in observation under microscope, and powder XRD matched with the pure phase, providing no decomposition for the vacuum-annealed sample.

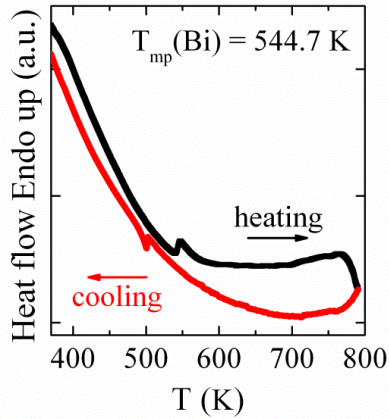

**Fig. 4** DTA plot for $BaMn_2Bi_2$ measured in flowing ultra-high-purity argon atmosphere. The cusp features correspond to the surface Bi melting (heating, a maximum) and freezing (cooling, a minimum).

In $Ba_{1-x}K_xMn_2Bi_2$, upon increasing the K content, the lattice parameter $a$ slightly increases to ~4.5 Å (Fig. 5a). The $c$-axis on the other hand decreases more noticeably down to ~14.60 Å, resulting in ~0.08 Å change. The changes in the $a$ and $c$ lattice parameters compensate to some extent and give a small shrinkage of 0.5 Å$^3$ by going from the parent to $x = 0.32$; these findings corroborate with virtually identical cationic radii of $Ba^{2+}$ (1.35 Å) and $K^+$ (1.38 Å).[33] For $Ba_{1-x}K_xFe_2As_2$, $a$ decreases and $c$ increases with increasing $x$,[5,34] giving approximately no change in cell volume. However, similarly, $c$ decreases with K-doping in $Ba_{1-x}K_xMn_2As_2$, which has been related to the strong *p-d* hybridization in BaMn$_2$As$_2$.[9,22] Following the same line of argument, the observed change of lattice parameters may point to a strong hybridization between Bi *p* and Mn *d* orbitals in BaMn$_2$Bi$_2$.

A representative EDS spectrum for the Ba$_{1-x}$K$_x$Mn$_2$Bi$_2$ solid solution is provided in Fig. 5b; the inset shows the intergrown plates of $x = 0.1$ crystals, measuring up to 1 cm. The K content was lower than the nominal compositions, with $x = 0.10(1)$ for nominal $x = 0.4$, $x = 0.32(1)$ for nominal $x = 0.6$, and $x = 0.36(1)$ for nominal $x = 0.8$. The loss of potassium during reactions is not unusual for the relatively high temperature flux growth of crystals containing alkali metals.[35,36]

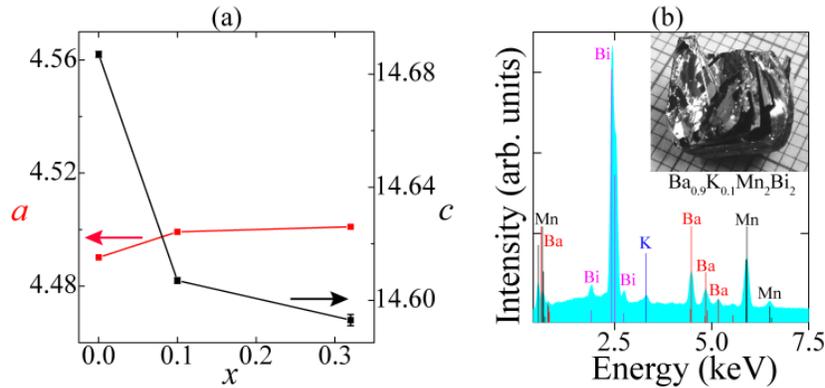

**Fig. 5** (a) Plot of the unit cell parameters versus measured *x* for Ba$_{1-x}$K$_x$Mn$_2$Bi$_2$. (b) EDS spectrum for Ba$_{1-x}$K$_x$Mn$_2$Bi$_2$ with $x = 0.10(1)$ crystal; inset is the typical picture for the as-grown $x = 0.10(1)$ crystals, measuring up to 1 cm in length.

**Physical properties**

Fig. 6 presents plots of temperature dependence of electrical resistivity in the *ab* plane. Electrical resistivity data for BaMn$_2$Bi$_2$ (Fig. 6a) is reminiscent of that of BaMn$_2$As$_2$ where $\rho$ diminishes with decreasing temperature in the 400–90 K range, after which it changes slope and starts increasing in low *T* region. The observed $\rho(T)$ plot is indicative of a small gap semiconductor. A band gap of $E_g = 6$ meV can be derived from the Arrhenius fit ($\ln\rho = \ln\rho_0 + E_g/2k_BT$) in the 60-25 K region. The calculated band gap for BaMn$_2$Bi$_2$ is 9 times lower than that reported for BaMn$_2$As$_2$ ($E_g = 54$ meV),[8] which in turn is ~ 3 times lower than that reported for BaMn$_2$P$_2$ ($E_g = 146$ meV).[17] If the trend holds, whereby band gaps are 3 times lower between each consecutive members going down the group, $BaMn_2Sb_2$ should similarly be a small gap semiconductor with $E_g \approx 18$ meV. The calculated band gaps for $BaMn_2P_2$, $BaMn_2As_2$ and $BaMn_2Bi_2$ also follow chemical trends in the periodic table, according to which lighter electronegative elements P and As form compounds with larger gaps compared to heavier and more electropositive Bi. For $BaMn_2Bi_2$, temperature dependence of electrical resistivity under applied field of 8 Tesla results in 50% negative magnetoresistance, $100\times(\rho_{8\,T} - \rho_{0\,T})/\rho_{0\,T}$, at 2 K.

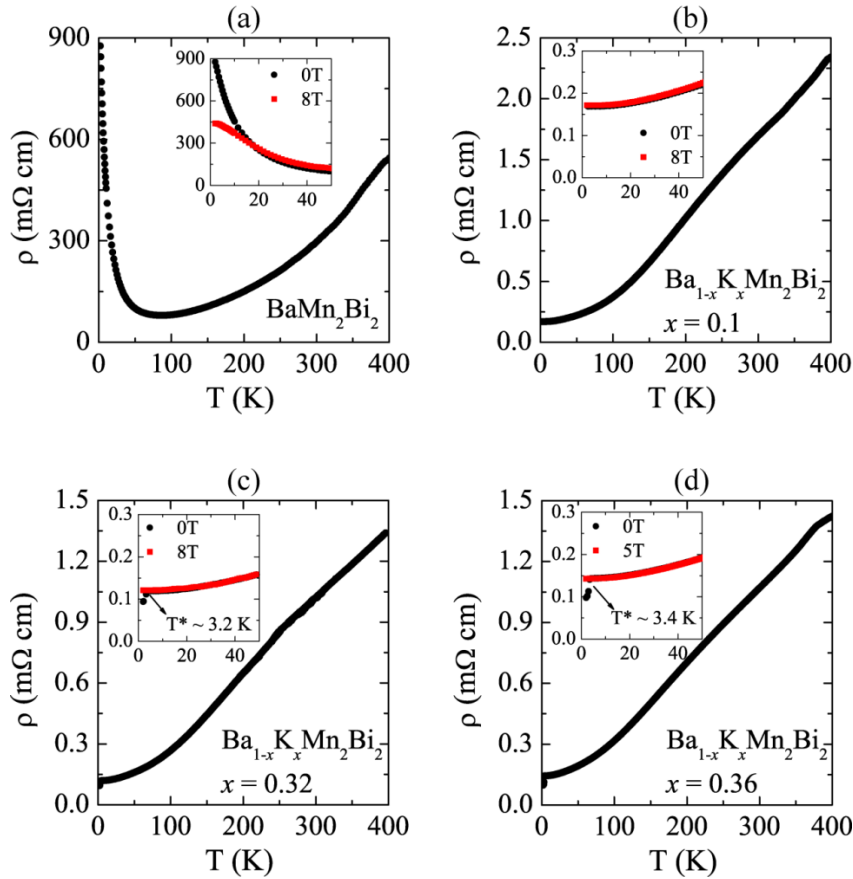

**Fig. 6** Temperature dependence of electrical resistivity, $\rho(T)$, in the *ab* plane for (a) $BaMn_2Bi_2$; for $Ba_{1-x}K_xMn_2Bi_2$ with (b) $x = 0.1$, (c) $x = 0.32$ and (d) $x = 0.36$. Insets illustrate magnetic field effects below 50 K. There is a downturn in $\rho(T)$ data for $x = 0.32$ and $x = 0.36$ at low temperatures, which disappears under high applied field of 8 Tesla.

Hole-doping via substitution of Ba with K turns material from semiconducting in $BaMn_2Bi_2$ to metallic behavior in $Ba_{1-x}K_xMn_2Bi_2$ (Fig. 6b-d). The room temperature electrical resistivity values are $\rho_{300\ K}$ ~ 300 mΩ cm for BaMn$_2$Bi$_2$, ~ 2 mΩ cm for $x$ = 0.1 and ~ 1 mΩ cm for both $x$ = 0.32 and $x$ = 0.36. The crystals used for resistivity measurements are of good quality, as indicated by the residual resistivity ratios (RRR = $\rho_{300\ K}/\rho_{4\ K}$) of 10, 8.4 and 7.5 for $x$ = 0.1, $x$ = 0.32 and $x$ = 0.36, respectively. There is a downturn ($T^*$) in the $\rho(T)$ data below 4 K for $x$ = 0.32 and $x$ = 0.36 (Fig. 6c, d), but it does not drop to zero and therefore, claim of bulk superconductivity cannot be made.

There is no magnetoresistance for $x$ = 0.1 (Fig. 6b, inset). The applied fields of 5 to 8 T for $x$ = 0.32 and $x$ = 0.36 shifts $T^*$ to lower temperatures, or eliminates it (Fig. 6c, 6d, insets). The warm up and cool down $\rho(T)$ data overlap, and no thermal hysteresis is observed, indicating that $T^*$ is not a first order transition. For the higher $x$ = 0.36, $T^*$ is sharper and at a slightly higher temperature of 3.4 K compared to $x$ = 0.32, suggesting composition dependence of $T^*$. Filamentary superconductivity may exist in these samples, origin of which may be possible strain or nanoscale phase segregation.[37] Strain-induced superconductivity, for example, has been reported for the parents of SrFe$_2$As$_2$[38] and BaFe$_2$As$_2$.[39] A field- and composition-dependent downturn in $\rho(T)$ was also reported for lightly Mo-doped BaFe$_2$As$_2$ crystals.[37]

Similar to BaMn$_2$Bi$_2$, BaMn$_2$As$_2$ also changes from semiconducting behavior to metallic upon K-doping,[21,22] but with no downturn in the $\rho(T)$ data. Metallization in BaMn$_2$As$_2$ with a downturn in resistivity below ~ 20 K can be induced by applied pressure of 8.2 GPa;[20,22] it was speculated that this may be fractional superconductivity and that higher pressures may give bulk superconductivity.[20] For Ba$_{1-x}$K$_x$Mn$_2$Bi$_2$, higher hole-doping levels and high-pressure experiments may shed light on the nature of $T^*$, and if it may be linked to a superconducting state.

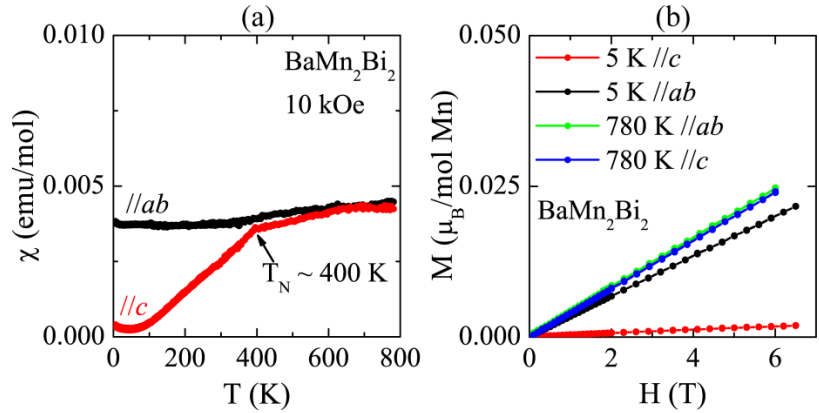

**Fig. 7** For BaMn$_2$Bi$_2$, anisotropic (a) temperature dependence of magnetic susceptibility and (b) field dependence of magnetization at 5 K and 780 K.

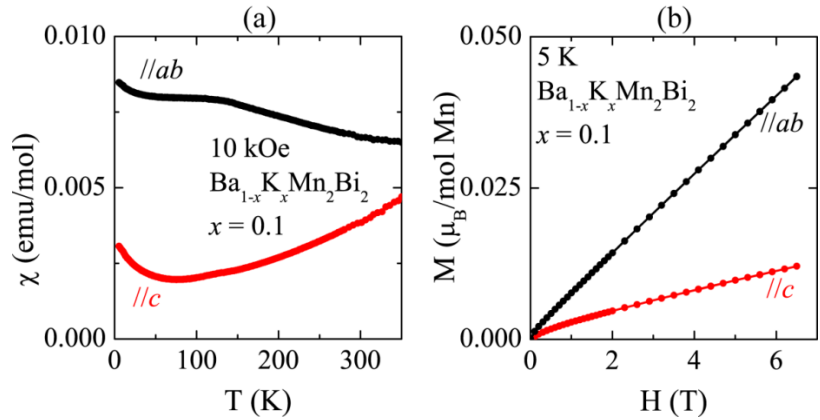

**Fig. 8** For Ba$_{1-x}$K$_x$Mn$_2$Bi$_2$ with $x$ = 0.10(1), anisotropic (a) temperature dependence of magnetic susceptibility and (b) field dependence of magnetization at 5 K.

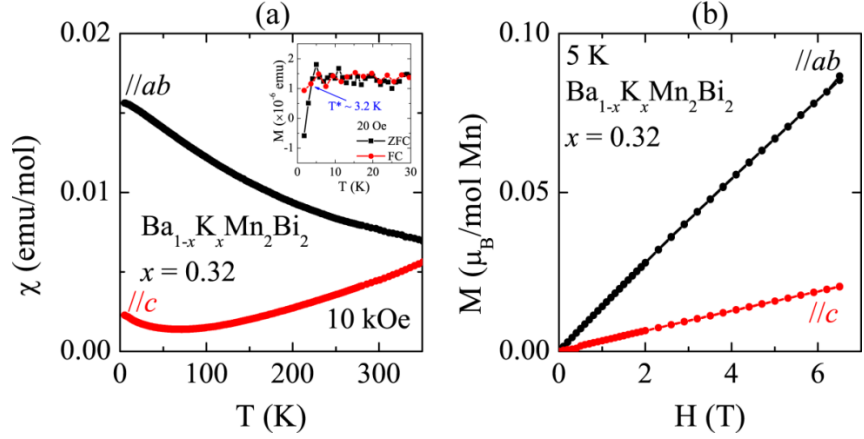

**Fig. 9** For Ba$_{1-x}$K$_x$Mn$_2$Bi$_2$ with $x = 0.32(1)$, anisotropic (a) temperature dependence of magnetic susceptibility and (b) field dependence of magnetization at 5 K.

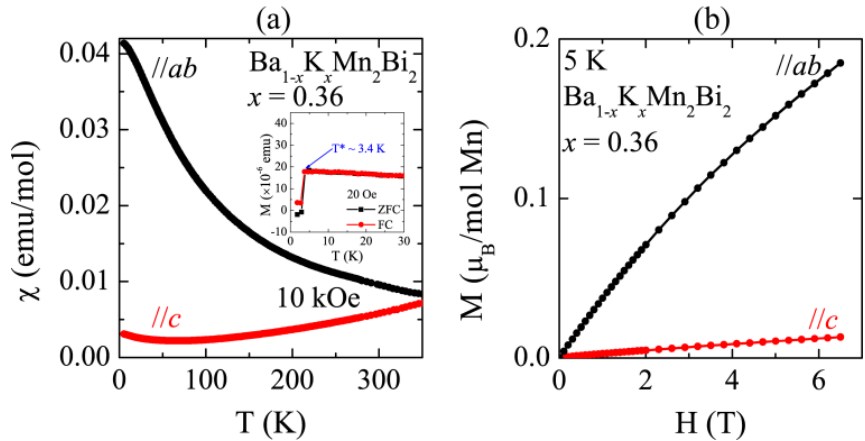

**Fig. 10** For Ba$_{1-x}$K$_x$Mn$_2$Bi$_2$ with $x = 0.36(1)$, anisotropic (a) temperature dependence of magnetic susceptibility and (b) field dependence of magnetization at 5 K.

Temperature and field dependence of anisotropic magnetic susceptibility and magnetization results for BaMn$_2$Bi$_2$ and Ba$_{1-x}$K$_x$Mn$_2$Bi$_2$ are plotted in Fig. 7 to 10; all of the ZFC and FC data overlap, thus only ZFC are shown. Higher amount of Bi flux and flux removal temperature of 500°C, which is above the temperature of peritectic formation of MnBi ($T_{per} = 446$°C)[40] yielded crystals free of surface ferromagnetic MnBi impurity (see Experimental section). For these pure crystals (Fig. 7), $M(H)$ plots (Fig. 7b) are linear at 5 K and 780 K. $\chi_{ab}(T)$ is temperature independent up to 400 K, after which it increases weakly; $\chi_c(T)$ also has a weak temperature dependence above 400 K where it slightly increases in values. However, at 400 K there is a discontinuity in $\chi_c(T)$, and it decreases linearly upon cooling down to 80 K and then plateaus close to zero. $\chi_c(T)$ plateaus presumably due to a paramagnetic orbital susceptibility.[8] The results indicate a noticeable anisotropy with $\chi_{ab}/\chi_c = 10$ at 5 K, and are in agreement with the reported $\chi_{ab}(T)$ and $\chi_c(T)$ for BaMn$_2$As$_2$, which has a comparable in and out of plane anisotropy, with $\chi_{ab}/\chi_c = 7.5$ at 2 K.[8] This anisotropic $\chi(T)$ suggest collinear antiferromagnetic order below $T_N \sim 400$ K. The fact that BaMn$_2$Bi$_2$ does not obey the Curie-Weiss law above the transition temperature is indicative of strong exchange coupling, and in this regard, the material is similar to compounds such as BaFe$_2$As$_2$,[6] LaFeAsO[41] and LaMnPO.[42]

The magnetization measurements on larger crystals of BaMn$_2$Bi$_2$, obtained by removing the flux at 415°C, are shown in Supplementary Information section (S1). $M(H)$ plots show hysteresis suggesting presence of a ferromagnetic impurity with saturation magnetization values of $M_{sat,ab} = 10.8 \times 10^{-3}$ μ$_B$/mol Mn and $M_{sat,c} = 10.5 \times 10^{-3}$ μ$_B$/mol Mn. If the presence of ferromagnetic MnBi ($T_c = 633$ K) is assumed, the above-mentioned $M_{sat}$ correspond to 0.24% MnBi surface impurity, considering $M_{sat} = 4.5(2)$ μ$_B$/mol Mn value for MnBi from neutron diffraction data.[43] A similar set of $M(H)$ plots has been obtained for BaMn$_2$As$_2$ with the proposed origin of the ferromagnetic component being MnAs,[8] which is seen in $\chi(T)$ data as a ferromagnetic transition at ~310 K.

K-doping in BaMn$_2$Bi$_2$ induces changes in magnetic behavior as evidenced in Fig. 8 to 10. Although there is an increase in numerical susceptibility values, the behavior of $\chi_c(T)$ remains almost unchanged at all doping levels, whereas $\chi_{ab}(T)$ is different. For $x = 0.1$, $\chi_{ab}(T)$ slightly increases with decreasing temperature. The increase in $\chi_{ab}(T)$ with decreasing temperature is more pronounced for higher doping levels of $x = 0.32$, and especially $x = 0.36$; the divergence between $\chi_{ab}(T)$ and $\chi_c(T)$ increases with rising K-content. The fact that anisotropy along and perpendicular to the [Mn$_2$Bi$_2$] layers increases with hole-doping is supported by $\chi_{ab}/\chi_c$ values at 5 K of ~ 3, 7 and 13 for $x = 0.1$, $x = 0.32$ and $x = 0.36$, respectively. Similarly, heavily doped Ba$_{1-x}$K$_x$Mn$_2$As$_2$ samples also demonstrate increased $\chi_{ab}/\chi_c$ values up to 15-25.[22] $M(H)$ plots at 5 K (Fig. 8b to 10b) are approximately linear for K-doped samples of $x = 0.1$ and $x = 0.32$. For $x = 0.36$, $M_c$ is linear, but there is an overall curvature in $M_{ab}$ plot. The linear $M(H)$ plots (see Supporting Information for $M(H)$ data at 295 K) suggest that the changes in $\chi_{ab}(T)$ and $\chi_c(T)$ cannot be attributed to ferromagnetic MnBi impurity. It can be speculated that although K-doped samples are quenched from 415°C, the formation of Mn-rich secondary phase (see Experimental section) prevents formation of MnBi. Magnetic measurements under 20 Oe were also carried out for $Ba_{1-x}K_xMn_2Bi_2$ in order to check for anomalies at $T^*$ (Fig. 9a and 10a, insets). Although no diamagnetism is seen, there is a feature in $\chi_{ab}(T)$ at $T^* \sim 3.4$ K for $x = 0.36$ and $T^* \sim 3.2$ K for $x = 0.32$.

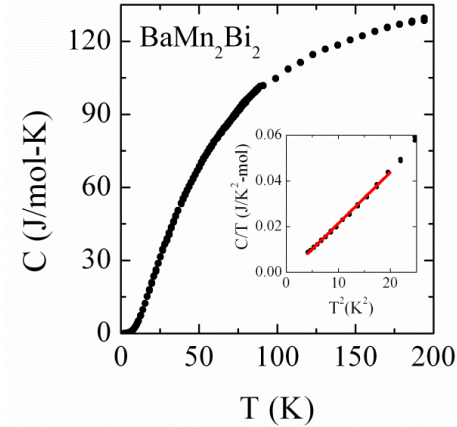

**Fig. 11** Heat capacity versus temperature, $C(T)$, for $BaMn_2Bi_2$ measured below 200 K. Inset shows $C(T)/T$ versus $T^2$ data below 5 K (black dots) and a linear fit by the expression $C/T = \gamma + \beta T^2$ (red line).

Temperature dependence of heat capacity for $BaMn_2Bi_2$ is shown in Fig. 11. Several scans below 200 K have been carried out with special attention to the 90-110 K regions where the broad feature in $\rho(T)$ was observed. No evidence of a phase transition is seen up to 200 K. At higher temperatures, the values of molar heat capacity reach 129 J/(K·mol). This value is close to the classical Dulong Petit heat capacity value given by $C = 3nR \sim 125$ J/(K·mol), where $n$ is the number of atoms per formula and $R$ is the molar gas constant. The obtained heat capacity values and absence of any phase transitions in this temperature interval are in accord with the reported heat capacity results for $BaMn_2As_2$.[8] From the linear fit of $C(T)/T$ vs $T^2$ (Fig. 11, inset), the electronic and lattice contributions to the heat capacity were estimated to be $\gamma \approx 0$ for the Sommerfeld coefficient and $\beta = 0.000434(3)$ J/(K$^2$ mol atom), respectively. The $\gamma$ value confirms insulating ground state suggested by the resistivity results. A Debye temperature of $\theta_D \approx 165$ K was calculated from the lattice heat capacity.

**Conclusions**

The first bismuthide BaMn$_2$Bi$_2$ with ThCr$_2$Si$_2$-type structure along with K-doped compositions of Ba$_{1-x}$K$_x$Mn$_2$Bi$_2$ are crystallized in large forms using bismuth flux. High solubility of Mn in Bi[40] can be utilized at lower temperatures, enabling hole-doping of BaMn$_2$Bi$_2$ with volatile K. Powder X-ray diffraction experiments give $a$ = 4.4902(3) Å, and $c$ = 14.687(1) Å for BaMn$_2$Bi$_2$. Upon substitution of Ba with K, $a$ lattice parameter remains almost unchanged, whereas $c$ slightly decreases. The reaction of ground BaMn$_2$Bi$_2$ with air results in a mixture containing Bi and BaBi$_3$. The latter binary is also air-sensitive, and powder XRD data of the sample left overnight in air gives only peaks from Bi and MnO$_2$.

Temperature dependence of electrical resistivity $\rho(T)$ for BaMn$_2$Bi$_2$ suggests a small gap semiconducting behavior. The calculated band gap in the low $T$ region is $E_g$ = 6 meV, a much smaller band gap compared to BaMn$_2$P$_2$ ($E_g$ = 146 meV)[17] and BaMn$_2$As$_2$ ($E_g$ = 54 meV).[8] The values of the band gaps in this family follow the periodic trend based on the electronegativities of corresponding pnictogen elements. The Sommerfeld coefficient of $\gamma \approx 0$ from heat capacity data is consistent with the insulating ground state. BaMn$_2$Bi$_2$ turns from a ground state insulator into a metal upon hole-doping via substitution of Ba with K in Ba$_{1-x}$K$_x$Mn$_2$Bi$_2$. At higher K-doping levels, there is a clear downturn at $T^*$ = 3.2 K and 3.4 K for $x$ = 0.32 and $x$ = 0.36, respectively, in both resistivity and magnetization. The nature of downturn is unclear, but can be tied to the traces of superconductivity, similar to that observed in strained SrFe$_2$As$_2$[38] and BaFe$_2$As$_2$.[39]

Results of $\chi(T)$ and $M(H)$ give collinear antiferromagnetic order for BaMn$_2$Bi$_2$. Below the ordering temperature of T$_N$ ~ 400 K, $\chi_c(T)$ goes to zero when cooled, and therefore, $c$ is the easy-axis direction; $\chi_{ab}(T)$ is temperature independent, indicating that $ab$ plane is perpendicular to the easy-axis. Above $T_N$ ~ 400 K, $\chi(T)$ mildly increases, which is indicative of a strong exchange coupling in BaMn$_2$Bi$_2$. The antiferromagnetic structure of BaMn$_2$Bi$_2$ is likely to be of $G$-type considering the fact that all other BaMn$_2$$Pn$$_2$ ($Pn$ = P, As, Sb)[9,17,18] pnictides order in this magnetic structure. Magnetization results for Ba$_{1-x}$K$_x$Mn$_2$Bi$_2$ are similar to those reported for Ba$_{1-x}$K$_x$Mn$_2$As$_2$ series.[8,22] At low doping levels, Ba$_{1-x}$K$_x$Mn$_2$As$_2$ ($x$ = 0.016 and $x$ = 0.05) were found to be antiferromagnetic local moment metals,[21] and the same likely applies to Ba$_{1-x}$K$_x$Mn$_2$Bi$_2$ ($x$ = 0.10), too.

In conclusion, the new BaMn$_2$Bi$_2$ (ThCr$_2$Si$_2$-type) compound is a local moment antiferromagnetic insulator, which turns metallic in $Ba_{1-x}K_xMn_2Bi_2$ with $0 < x \leq 0.36$. Neutron experiments are necessary on $Ba_{1-x}K_xMn_2Bi_2$ to confirm antiferromagnetic behavior of these materials. The downturns in $\rho(T)$ and $\chi(T)$ data for the two K-doped samples may suggest the possibility of the onset of superconductivity, so higher K-doped samples and pressure experiments can be explored. The new $BaMn_2Bi_2$ compound reported here may potentially be a new parent superconductor with $ThCr_2Si_2$-type structure. This work shows a controlled variation of magnetic and electrical properties of a new material via chemical synthesis and doping, which may be useful for preparation and improvement of different types of functional materials.

## Acknowledgments

This work was supported by the Department of Energy, Basic Energy Sciences, Materials Sciences and Engineering Division. We thank R. Custelcean for his assistance with the single crystal X-ray diffraction experiments.

## References

1   R. Hoffmann and C. Zheng, *J. Phys. Chem.*, 1985, **89**, 4175.
2   F. Steglich, J. Aarts, C. D. Bredl, W. Lieke, D. Meschede, W. Franz and H. Schifer, *Phys. Rev. Lett.*, 1979, **43**, 1892.
3   S. A. Jia, P. Jiramongkolchai, M. R. Suchomel, B. H. Toby, J. G. Checkelsky, N. P. Ong and R. J. Cava, *Nature Phys.*, 2011, **7**, 207.
4   M. Rotter, M. Tegel and D. Johrendt, *Phys. Rev. Lett.*, 2008, **101**, 107006.
5   M. Rotter, M. Pangerl, M. Tegel and D. Johrendt, *Angew. Chem. Int. Ed.*, 2008, **47**, 7949.
6   A. S. Sefat, R. Y. Jin, M. A. McGuire, B. C. Sales, D. J. Singh and D. Mandrus, *Phys. Rev. Lett.*, 2008, **101**, 117004.
7   D. J. Singh, A. S. Sefat, M. A. McGuire, B. C. Sales, D. Mandrus, L. H. VanBebber and V. Keppens, *Phys. Rev. B*, 2009, **79**, 094429.
8   Y. Singh, A. Ellern and D. C. Johnston, *Phys. Rev. B*, 2009, **79**, 094519.
9   J. An, A. S. Sefat, D. J. Singh and M. H. Du, *Phys. Rev. B*, 2009, **79**, 075120.
10  B. Saparov and A. S. Sefat, *J. Solid State Chem.*, 2012, **191**, 213.
11  A. S. Sefat, D. J. Singh, R. Jin, M. A. McGuire, B. C. Sales and D. Mandrus, *Phys. Rev. B*, 2009, **79**, 024512.

12   A. S. Sefat, D. J. Singh, R. Jin, M. A. McGuire, B. C. Sales, F. Ronning and D. Mandrus, *Physica C*, 2009, **469**, 350.

13   L. Zhang, A. Subedi, D. J. Singh and M. H. Du, *Phys. Rev. B*, 2008, **78**, 174520.

14   X. C. Wang, Q. Q. Liu, Y. X. Lv, W. B. Gao, L. X. Yang, R. C. Yu, F. Y. Li and C. Q. Jin, *Solid State Comm.*, 2008, **148**, 538.

15   *ICSDWeb*, Version 2.1.1, FIZ Karlsruhe, 2011.

16   E. Brechtel, G. Cordier and H. Schäfer, *Z. Naturforsch., Teil B*, 1979, **34**, 921.

17   S. L. Brock, J. E. Greedan and S. M. Kauzlarich, *J. Solid State Chem.* 1994, **113**, 303.

18   Y. Singh, M. A. Green, Q. Huang, A. Kreyssig, R. J. McQueeney, D. C. Johnston and A. I. Goldman, *Phys. Rev. B*, 2009, **80**, 100403.

19   A. Antal, T. Knoblauch, Y. Singh, P. Gegenwart, D. Wu and M. Dressel, *Phys. Rev. B*, 2012, **86**, 014506.

20   A. T. Satya, A. Mani, A. Arulraj, N. V. Chandra Shekar, K. Vinod, C. S. Sundar and A. Bharathi, *Phys. Rev. B*, 2011, **84**, 180515.

21   A. Pandey, R. S. Dhaka, J. Lamsal, Y. Lee, V. K. Anand, A. Kreyssig, T. W. Heitmann, R. J. McQueeney, A. I. Goldman, B. N. Harmon, A. Kaminsky and D. C. Johnston, *Phys. Rev. Lett.*, 2012, **108**, 087005.

22   J. K. Bao, H. Jian, Y. L. Sun, W. H. Jiao, C. Y. Shen, H. J. Guo, Y. Chen, C. M. Feng, H. Q. Yuan, Z. A. Xu, G. H. Cao, R. Sasaki, T. Tanaka, K. Matsuyabashi and Y. Uwatoko, *Phys. Rev. B*, 2012, **85**, 144523.

23   S. Q. Xia, C. Myers and S. Bobev, *Eur. J. Inorg. Chem.*, 2008, **2008**, 4262.

24   H. F. Wang, K. F. Cai, H. Li, L. Wang and C. W. Zhou, *J. Alloys Compd.*, 2009, **477**, 519.

25   A. C. Larson and R. B. Von Dreele, *General Structure Analysis System (GSAS)*, Los Alamos National Laboratory Report LAUR 86-748, 2000.

26   B. H. Toby, *J. Appl. Crystallogr.*, 2001, **34**, 210.

27   R. Marchand and W. Jeitschko, *J. Solid State Chem.*, 1978, **24**, 351.

28   S. Ran, S. L. Bud'ko, D. K. Pratt, A. Kreyssig, M. G. Kim, M. J. Kramer, D. H. Ryan, W. N. Rowan-Weetaluktuk, Y. Furukawa, B. Roy, A. I. Goldman and P. C. Canfield, *Phys. Rev. B*, 2011, **83**, 144517.

29   P. Villars and K. Cenzual, *Pearson's Crystal Data: Crystal Structure Database for Inorganic Compounds,* ASM International, Materials Park, Ohio, 2008/9.

30   T. Y. Kuromoto, S. M. Kauzlarich, and D. J. Webb, *Chem. Mater.*, 1992, **4**, 435.

31   H. U. Schuster and G. Achenbach, *Z. Naturforsch., Teil B*, 1978, **33**, 113.

32   S. Ponou, T. F. Fässler and L. Kienle, *Angew. Chem. Int. Ed.*, 2008, **47**, 3999.

33  R. D. Shannon, *Acta Crystallogr., Sect. A: Cryst. Phys., Diffr., Theor. Gen. Crystallogr.*, 1976, **32**, 751.

34  H. Chen, Y. Ren, Y. Qiu, W. Bao, R. H. Liu, G. Wu, T. Wu, Y. L. Xie, X. F. Wang, Q. Huang and X. H. Chen, *EPL*, 2009, **85**, 17006.

35  H. Luo, Z. Wang, H. Yang, P. Cheng, X. Zhu and H. H. Wen, *Supercond. Sci. Technol.*, 2008, **21**, 125014.

36  N. Ni, S. L. Bud'ko, A. Kreyssig, S. Nandi, G. E. Rustan, A. I. Goldman, S. Gupta, J. D. Corbett, A. Kracher and P. C. Canfield, *Phys. Rev. B*, 2008, **78**, 014507.

37  A. S. Sefat, K. Marty, A. D. Christianson, B. Saparov, M. A. McGuire, M. D. Lumsden, W. Tian and B. C. Sales, *Phys. Rev. B*, 2012, **85**, 024503.

38  S. R. Saha, N. P. Butch, K. Kirshenbaum, J. Paglione and P. Y. Zavalij, *Phys. Rev. Lett.*, 2009, **103**, 037005.

39  M. A. Tanatar, N. Ni, G. D. Samolyuk, S. L. Bud'ko, P. C. Canfield and R. Prozorov, *Phys. Rev. B*, 2009, **79**, 134528.

40  *ASM Alloy Phase Diagrams Center*, P. Villars, editor-in-chief; H. Okamoto and K. Cenzual, section editors; http://www1.asminternational.org/AsmEnterprise/APD, ASM International, Materials Park, OH, 2006.

41  M. A. McGuire, A. D. Christianson, A. S. Sefat, B. C. Sales, M. D. Lumsden, R. Jin, E. A. Payzant, D. Mandrus, Y. Luan, V. Keppens, V. Varadarajan, J. W. Brill, R. P. Hermann, M. T. Sougrati, F. Grandjean and G. J. Long, *Phys. Rev. B*, 2008, **78**, 094517.

42  J. W. Simonson, Z. P. Yin, M. Pezzoli, J. Guo, J. Liu, K. Post, A. Efimenko, N. Hollmann, Z. Hu, H.-J. Lin, C.-T. Chen, C. Marques, V. Leyva, G. Smith, J. W. Lynn, L. L. Sun, G. Kotliar, D. N. Basov, L. H. Tjeng and M. C. Aronson, *Proc. Natl. Acad. Sci. USA*, 2012, **109**, E1815.

43  A. F. Andresen, W. Hälg, P. Fischer and E. Stoll, *Acta Chem. Scand.*, 1967, **21**, 1543.